\documentclass[a4paper,fleqn]{cas-dc}

\usepackage[round,authoryear]{natbib}
\usepackage{subfigure}
\usepackage{amsmath}
\usepackage[ruled]{algorithm2e}
\usepackage{soul}
\usepackage{bm}
\usepackage{color, xcolor}
\soulregister\cite7
\soulregister\citep7
\soulregister\ref7
 
\usepackage{tabularx}
\usepackage{changepage}
\usepackage[T1]{fontenc}
\usepackage[latin9]{inputenc}
\usepackage{xcolor}
\usepackage{array}
\usepackage{float}
\usepackage{amsthm}
\usepackage{amssymb}
\usepackage{multirow}
\usepackage{booktabs}
\usepackage{multirow}
\usepackage{graphicx}
\usepackage{makecell}
\usepackage{enumitem}
\usepackage{longtable}
\usepackage{rotating}
\usepackage{appendix}
\usepackage{soul}

\def\tsc#1{\csdef{#1}{\textsc{\lowercase{#1}}\xspace}}
\tsc{WGM}
\tsc{QE}
\definecolor{feature_type}{RGB}{230, 243, 248}
\definecolor{view_type}{RGB}{234, 236, 181}
\definecolor{feature_subtype}{RGB}{206, 204, 248}
\begin{document}
\let\WriteBookmarks\relax
\def\floatpagepagefraction{1}
\def\textpagefraction{.001}
\let\printorcid\relax

\shorttitle{w. Zhao et~al.}
\shortauthors{zwx et~al.}

% Main title of the paper
\title [mode = title]{AppPoet: Large Language Model based Android malware detection via multi-view prompt engineering}  

% Title footnote mark
% eg: \tnotemark[1]
% \tnotemark[<tnote number>] 

% Title footnote 1.
% eg: \tnotetext[1]{Title footnote text}
% \tnotetext[<tnote number>]{<tnote text>} 

% First author
%
% Options: Use if required
% eg: \author[1,3]{Author Name}[type=editor,
%       style=chinese,
%       auid=000,
%       bioid=1,
%       prefix=Sir,
%       orcid=0000-0000-0000-0000,
%       facebook=<facebook id>,
%       twitter=<twitter id>,
%       linkedin=<linkedin id>,
%       gplus=<gplus id>]

\author[author1]{Wenxiang Zhao}
% Corresponding author indication

% Footnote of the first author
% \fnmark[College of Computer Science and Engineering, Chongqing University of Technology, Chongqing 400054, China]
% Email id of the first author
\ead{zhaowx98@ustc.edu.cn}
% URL of the first author
% \ead[url]{<URL>}
% Credit authorship
% eg: \credit{Conceptualization of this study, Methodology, Software}
% \credit{<Credit authorship details>}

% Address/affiliation
% \affiliation[<aff no>]{organization={},
%             addressline={}, 
%             city={},
% %          citysep={}, % Uncomment if no comma needed between city and postcode
%             postcode={}, 
%             state={},
%             country={}}

\author[author1]{Juntao Wu}   
\ead{wjt99@mail.ustc.edu.cn}
\author[author2]{Zhaoyi Meng}
\cormark[1]
\ead{zymeng@ahu.edu.cn}

\address[author1]{School of Management, University of Science and Technology of China, Hefei, China}
\address[author2]{School of Computer Science and Technology, Anhui University, Hefei, China}

% Corresponding author text
\cortext[1]{Corresponding author.}

% Footnote text
% \fntext[1]{}

% For a title note without a number/mark
%\nonumnote{}

% Here goes the abstract
\begin{abstract}
Due to the vast array of Android applications, their multifarious functions and intricate behavioral semantics, attackers can adopt various tactics to conceal their genuine attack intentions within legitimate functions. 
However, numerous learning-based methods suffer from a limitation in mining behavioral semantic information, thus impeding the accuracy and efficiency of Android malware detection. 
Besides, the majority of existing learning-based methods are weakly interpretive and fail to furnish researchers with effective and readable detection reports. 
Inspired by the success of the Large Language Models (LLMs) in natural language understanding, we propose AppPoet, a LLM-assisted multi-view system for Android malware detection.
Firstly, AppPoet employs a static method to comprehensively collect application features and formulate various observation views. 
Then, using our carefully crafted multi-view prompt templates, it guides the LLM to generate function descriptions and behavioral summaries for each view, enabling deep semantic analysis of the views.
Finally, we collaboratively fuse the multi-view information to efficiently and accurately detect malware through a deep neural network (DNN) classifier and then generate the human-readable diagnostic reports. 
Experimental results demonstrate that our method achieves a detection accuracy of 97.15\% and an F1 score of 97.21\%, which is superior to the baseline methods.
Furthermore, the case study evaluates the effectiveness of our generated diagnostic reports.
\end{abstract}

% Use if graphical abstract is present
%\begin{graphicalabstract}
%\includegraphics{}
%\end{graphicalabstract}

\maketitle

%%%%%%%%%%%%%%%%%%%%%%%%
% \begin{table}[<options>]  
% \caption{}\label{tbl1}
% \begin{tabular*}{\tblwidth}{@{}LL@{}}
% \toprule
%   &  \\ % Table header row
% \midrule
%  & \\
%  & \\
%  & \\
%  & \\
% \bottomrule
% \end{tabular*}
% \end{table}

% Numbered list
% Use the style of numbering in square brackets.
% If nothing is used, default style will be taken.
%\begin{enumerate}[a)]
%\item 
%\item 
%\item 
%\end{enumerate}  

% Unnumbered list
%\begin{itemize}
%\item 
%\item 
%\item 
%\end{itemize}  

% Description list
%\begin{description}
%\item[]
%\item[] 
%\item[] 
%\end{description}  

\section{Introduction}

With the advancement of technology, mobile devices have become integral to people's daily lives. According to "The Mobile Economy 2023" reported by Global System for Mobile communications Association (GSMA) \citep{GSMA}, by the end of 2022, there were 5.4 billion unique mobile subscribers worldwide, with 4.4 billion using mobile internet. By 2030, this number is expected to rise to 6.3 billion for subscribers and 5.5 billion for mobile internet users. Among the range of mobile operating systems, Android has been particularly vulnerable to malware attacks due to its open-source nature. According to statista \citep{statistics}, in the third quarter of 2023, over 438,000 instances of mobile malware installation were detected, marking an approximately 19\% increase from the second quarter. The proliferation of malware gravely compromises the privacy, property, and personal safety of users, posing significant risks to social stability and national security. Furthermore, as application function continues to expand, malware increasingly seeks to conceal its malicious intent within seemingly legitimate features. Therefore, determining how to effectively detect it remains a persistently pressing issue.

To tackle the issue mentioned above, numerous detection approaches have been proposed. 
Among them, learning-based detection methods have attracted attention for their detection accuracy and generalization ability.
These methods extract features extensively based on static methods without executing applications.
After encoding and transforming the extracted features, a representation vector for each target application is generated, which is then used to train a classifier to distinguish malware from benign applications.
Based on how features are utilized, learning-based methods can be classified into three categories: String-based, Image-based, and Graph-based approaches.
String-based methods \citep{arp2014drebin, zhu2023android} primarily arrange the extracted features as sequences of strings, which are then encoded into machine-readable vectors.
While these methods are easy to understand and straightforward to implement, they often fail to capture the semantic relationships between features, resulting in decreased detection accuracy.
Image-based methods \citep{sun2021android, tang2024android} convert APKs into images and apply image recognition techniques for classification.
Despite their simplicity and high efficiency, these methods tend to overlook critical semantic information within apps, leading to a reduction in accuracy. 
Moreover, the use of black-box models complicates result interpretation, limiting the generation of human-readable insights.
Graph-based methods \citep{onwuzurike2019mamadroid,wu2019malscan,hou2021disentangled} construct graph structures to capture the semantic relationships between features. 
Although they can more effectively represent complex application behaviors, constructing large or intricate graph structures introduces challenges related to computational efficiency and resource consumption.
Additionally, none of these approaches excel at producing human-readable and insightful reports, making it difficult for security experts to audit and analyze the results effectively.

Large Language Models (LLMs) have recently gained attention for their ability to excel in a wide range of tasks, from natural language understanding to complex reasoning. 
For example, OpenAI's GPT-3.5 \citep{openai}, trained on massive data resources and with 175 billion parameters, can easily perform a variety of tasks such as text generation \citep{gao2023enabling}, language translation \citep{wang2023document}, program code generation \citep{jiang2023self}, etc. 
These models have proven to be highly versatile, functioning across numerous domains as knowledgeable experts.
Leveraging these strengths, we sought to address the limitations of traditional learning-based detection methods by employing LLMs to extract and interpret the semantic relationships within application features.
Inspired by LLM's prompt engineering \citep{liu2023pre}, we designed structured and precise prompt workflows tailored to the characteristics of Android applications.
By utilizing LLM to act as Android security analysts, our approach enables the model to analyze feature string sequences, summarize the functions of features, and infer their potential behaviors.
Compared to String-based and Image-based methods, our approach leverages LLM to conduct deep semantic analysis of the extracted features. This allows us to capture not only the explicit function meanings but also the implicit relationships between features, thereby improving the accuracy of the analysis. In contrast to Graph-based methods, which can be computationally intensive and difficult to scale, our approach is more efficient and scalable while still maintaining robust interpretability.

In this work, we developed AppPoet, an LLM-assisted system for detecting Android malware and generating diagnostic reports. 
First, AppPoet selects typical features (including permission, API, URL, and uses-feature) accumulated by traditional String-based methods \citep{arp2014drebin}, and classifies them into different observation views according to the type and information content of the features further. 
Based on the different types of views, the LLM, with its rich knowledge, is used to generalize the functions and potential behaviors of the features within each view.
Subsequently, the pre-trained embedding model is utilized to convert all the textual information into machine-readable representation vectors, which are then fed into the trained DNN classifier \citep{schmidhuber2015deep} in a multi-view fusion manner to obtain the detection results. 
Finally, the LLM is guided to review all known information to generate a readable, valid diagnostic report.

However, a major challenge for AppPoet is to enable LLM to comprehend the features of different views, while outputting the factually correct feature function and inference summary based on the expert knowledge. 
Specifically, although LLM exhibits strong performance in natural language understanding and logical reasoning, if the task is vaguely defined or overly complex, LLM might process inputs and produce outputs with errors or fabricated content, i.e., LLM's "hallucination" \citep{zhang2023siren}. 

To address the above challenge, in this work, we propose the multi-view prompt engineering approach. 
First, we decompose the task into two phases: function description generation and view summary generation. In the first phase, we provide representative examples to facilitate LLM in generating function descriptions that meet the requirements through in-context learning \citep{brown2020language,xie2021explanation,workrethinking}. 
Based on the function description lists, we supply the LLM with detailed steps, requirements, and relevant terminology for generating view summaries using chain-of-thought \citep{wei2022chain,kojima2022large,wang2022self} reasoning. 
This ensures that the LLM fully understands the task and generates summaries with the same cognitive process.
To handle multiple views, we carefully designed function and summary templates, allowing the LLM to sequentially generate descriptions and summaries by incorporating feature information from various views.
It is worth noting that to further mitigate the negative effects of LLM's "hallucination," rather than relying solely on LLM to ascertain the maliciousness of a target application based on the known information, our method implements a model cascade approach to train a classifier utilizing a large volume of real samples for the discrimination task. 

To assess the performance of AppPoet, we conduct a comprehensive evaluation.
First, we collect 11,189 real benign apps and 12,128 malicious apps from AndroZoo \citep{allix2016androzoo} for training and testing, and our approach achieves higher accuracy, i.e., 97.15\% detection accuracy and 97.21\% F1 value, compared to the representative learning-based baseline methods \citep{arp2014drebin,onwuzurike2019mamadroid,wu2019malscan}. 
Second, we evaluate the effectiveness of the multi-view prompt engineering approach used by AppPoet through a comprehensive ablation experiment. 
In addition, we demonstrate the ability of our fuction memory component to improve efficiency and reduce cost through a comparison experiment. 
Finally, we evaluated the instructive value and significance of the diagnostic reports generated by AppPoet through a case. 
The contributions of this paper are as follows:

\begin{itemize}[itemsep=\parskip]

\item To the best of our knowledge, our work is an initial exploration of employing LLM for the task of Android malware detection. 
Based on the powerful inference and summarization capabilities of LLM, we intuitively generalize the explicit behavioral semantic information among application features to further releasing their detection potential and interpretability.

\item Our proposed multi-view prompt engineering approach significantly enhances the quality and stability of LLM output. 
Meanwhile, the detection accuracy and generalization ability are enhanced by the collaborative fusion of multi-view information.

\item Experiments indicate that our approach outperforms the existing typical baseline methods. 
Besides, we validate the effectiveness of the diagnostic reports through a case study.
\end{itemize}

\section{Related work}
\label{sec:2}

\subsection{Learning-based Android malware detection}

Learning-based Android malware detection methods accomplish detection task mainly through machine learning and deep learning techniques.
This type of methods usually starts by decompiling the APK.
Next, a variety of different features are extracted and selected, and are combined and processed in different ways to obtain the applied representation vector.
Finally, these representation vectors are used to train a classifier to detect malware.
We categorize learning-based methods into String-based, Image-based, and Graph-based in terms of the different uses of features and introduce each of them separately.

\subsubsection{String-based detection methods}

String-based methods typically organize features into a sequence of strings, which are then encoded into machine-readable vectors for training classifiers.
Different researchers have selected a wide variety of features to construct sequences from different perspectives.
Some works \citep{li2018significant,arslan2019permission,csahin2023novel} get permissions declared by apps from the AndroidManifest.xml files and compose sequences with different feature selection methods.
APIs are also one of the key features of interest to the research community. AppContext \citep{yang2015appcontext} leverage static analysis to extract contextual features of sensitive APIs in apps, including events that trigger sensitive APIs and control factors related to sensitive APIs. 
\cite{yumlembam2023android} captures the difference in usage between benign and malware APIs by introducing the BM25 (Best Matching 25) scoring feature, which calculates the BM25 score for each API. 
In addition to focusing on one key feature to construct a sequence, more methods \citep{arp2014drebin,kim2018multimodal,dhalaria2020framework,qiu2022cyber,cai2021jowmdroid,zhu2023android} use a wide range of features to construct representation sequences.
The most representative is Drebin \citep{arp2014drebin}, which extensively extracted kinds of features within apps, including permissions, APIs, hardware, and network.
String-based methods often employ feature engineering to select or refine features, but they lack sufficient depth in semantic analysis. 
This limitation makes it difficult to recognize contextual associations and interactions among features to uncover potential malicious behavior patterns, which adversely affects detection accuracy.

\subsubsection{Image-based detection methods}

Image-based methods consider how bytecode can be converted into an image and then detected using image recognition algorithms.
These methods \citep{hsien2018r2,xiao2019image} typically map bytecode to each channel of RGB, thus converting .dex file to an RGB image, and then use these RGB-encoded representation vectors to train CNN classifiers.
For better detection, recent work has used more complex and advanced classification algorithms instead of CNNs. \cite{zhu2023effective} selects vital parts of .dex file to be described as an RGB image and then uses the proposed novel CNN variant classifier for detection. \cite{tang2024android} proposes a method based on novel hybrid bytecode image and deep neural network combined with attention mechanism.
\cite{sun2021android} combine .dex file, .so file, and .xml file by mapping them to different RGB channels to create a image for detection.
It should be noted that such methods tend to use an end-to-end architecture. 
They can be rapidly used for detection after obtaining features with only a relatively shallow-level of processing.
It means that these methods have a huge advantage in terms of detection efficiency, but they ignore critical semantic information in apps thus causing a loss of accuracy.
In addition, Image-based methods are often regarded as black-box models that are difficult to interpret.

\subsubsection{Graph-based detection methods}

Different from processing features as a sequence of strings or converting them to image, Graph-based methods use the extracted features to construct graph structures that contain various semantic information. 
MaMaDroid \citep{onwuzurike2019mamadroid} statically extracts API calls from APKs and abstracts them into the form of family calls or package calls, and then models the feature vectors for training the classification model through Markov chains.
Malscan \citep{wu2019malscan} constructs function calls graphs from smali file of APKs and selects the sensitive API calls from it based on PScout.
CDGDroid \citep{xu2018cdgdroid} uses control-flow graphs, data-flow graphs and their possible combinations as features to characterize APKs.
AMCDroid \citep{liu2023enhancing} models application behavior as a homogeneous graph based on call graphs and code statements.
\cite{chen2024android} proposed a new type of call graph called the class-set call graph (CSCG), which takes Java class sets as nodes and call relationships between class sets as edges. 
These methods then use different graph representation learning methods to transform these graphs into representation vectors to train classification models for detection. 
In addition, there also some works \citep{hei2021hawk,hou2021disentangled,ye2019out} extract a wide range of features from APKs to construct heterogeneous information networks for malware detection.
There is no doubt that Graph-based methods are far more capable of mining semantic information for apps than String-based and Graph-based methods.
However, they still have the following limitations. 
On one hand, constructing a graph that is sufficient to adequately represent semantic information consumes lots of resources, especially for methods based on static analysis to obtain information-flow graphs.
On the other hand, while graph-based methods implicitly mine the semantics of application behaviors through graph representations, they often struggle to provide intuitive insights for human experts to conduct fine-grained auditing and analysis.

\subsection{Large language model}

Pre-trained Large Language Models (LLMs) \citep{chowdhery2023palm,zhang2022opt,ouyang2022training,touvron2023llama,vaswani2017attention} represented by ChatGPT has opened a new era of natural language understanding, which has been trained on a mega corpus and can support a wide range of natural language processing tasks through prompt engineering \citep{liu2023pre,chen2023unleashing,zhou2022large}. 
Compared to NLP with manual design patterns and ML with massive amounts of training data, prompt engineering is extremely lightweight. 
By simply describing natural language prompts, LLM can be invoked to perform specific tasks without additional training or hard-coding. 
To further unleash the great potential of LLM, the research community is constantly optimizing prompt engineering methods, such as in-context learning \citep{brown2020language,xie2021explanation,workrethinking}, and chain-of-thought prompting \citep{wei2022chain,kojima2022large,wang2022self}.

Due to its powerful text understanding and reasoning capabilities, researchers have employed LLM to solve various tasks within the Android domain \citep{feng2024prompting,liu2023make,liu2024testing,huang2024crashtranslator}. 
To the best of our knowledge, there is still no work that directly uses LLM for Android malware detection, but inspired by LLM's potential to understand behaviors, the idea of guiding LLM's inference and summarization through a prompt engineering approach is well suited to be applied to the identification of malicious behaviors.

\section{System architecture}
\label{sec:3}

The system architecture of AppPoet is illustrated in Figure \ref{fig:1}, which is developed for Android malware detection. It comprises the subsequent four modules: 

\begin{figure*}[!t]
\centering
\includegraphics[width=1.0\linewidth]{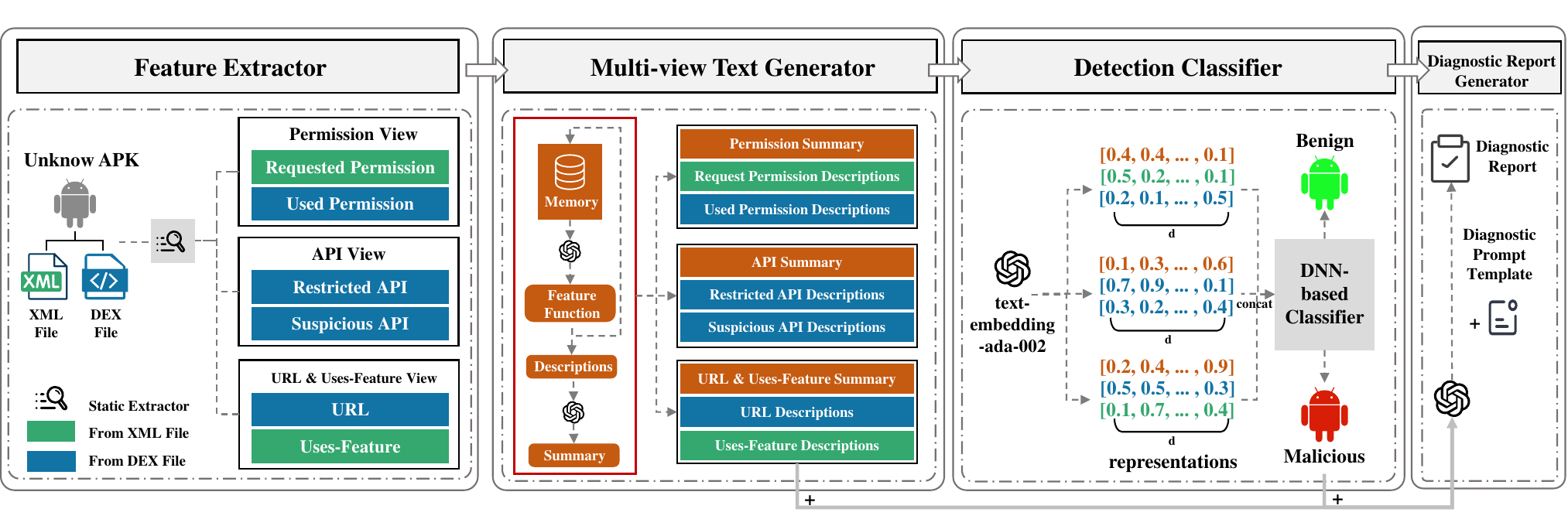}
\caption{System architecture of AppPoet.}
\label{fig:1}
\end{figure*}

\begin{itemize}[itemsep=\parskip]
\item \textbf{Feature extractor.} In this module, a feature extractor is developed based on static analysis. 
This extractor decompiles a given APK file and autonomously extracts selected features from the APK's \textit{AndroidManifest.xml} file and \textit{class.dex} file. 
The features primarily derive from permission, API, URL, and uses-feature. 
To facilitate unified modeling for subsequent modules, URL and uses-feature are merged, and the aforementioned features are categorized into three views: \textit{Permission View}, \textit{API View}, and \textit{URL \& uses-feature View}. (Refer to Section \ref{sec:4.1} for further details.)

\item \textbf{Multi-view text generator.} Building on the features extracted from the three views in the previous module, we propose a novel multi-view prompt engineering method. This approach aims to guide the LLM in generating descriptions and summaries across different views. To achieve this, we design \textit{function description prompt template} and \textit{view summary prompt template}, providing a unified framework that allows the LLM to generate standardized texts for a given APK. (Refer to Section \ref{sec:4.2} for further details.)

\item \textbf{Detection classifier.} Given the texts (descriptions and summaries) from different views, this module transforms all the texts into the machine-readable representation vectors, which are then concatenated into a single representation vector for describing the behavioral semantic information of the APK. 
Then, a DNN-based classifier is developed to learn the potential importance of the representations and give its prediction (i.e., a given unknown APK will be predicted to be malicious or not). (Refer to Section \ref{sec:4.3} for further details.)

\item \textbf{Diagnostic report generator.} To provide a more intuitive understanding of a given APK's potential malicious behavior, this module goes beyond a simple binary result (malicious or benign). It combines the descriptions and summaries from different views, along with the detection results, into a specially designed \textit{diagnostic report prompt template}. 
The LLM is then employed to generate a diagnostic report for the given APK, which offers preliminary insight into potential behaviors and provides a foundation for further exploration and validation. (Refer to Section \ref{sec:4.4} for further details.)
\end{itemize}

\section{Proposed methodology}
\label{sec:4}

This section provides a comprehensive overview of how AppPoet extracts, utilizes, and integrates the features of APK files into different views. 
Then we detail the process of acquiring descriptions, summaries, and vector representations for each view through the use of LLM.
Finally, this section describes how AppPoet discriminates between malware and benign application, as well as illustrates the methodology for generating the readable diagnostic reports.

\subsection{Feature selection and extraction}
\label{sec:4.1}

To describe the behavioral semantic information of Android applications in a more comprehensive way, inspired by Drebin \citep{arp2014drebin}, a classical String-based work in Android malware detection, we select four main feature types, namely, permission, API, URL, and uses-feature. 
Referring to Drebin, we subdivide permission into \textit{requested permission} and \textit{used permission}, as well as subdivide API into \textit{restricted API} and \textit{suspicious API} to further characterize the relevant behaviors. 
These features are then organized in the form of views to further model the behavioral semantics between them. 
Note that since a large number of applications do not hardcode URL and declare uses-feature, their number and frequency are much smaller than that of permission and API. Therefore, dividing them into separate views leads to unnecessary resource consumption and also renders a highly unbalanced information content expressed between the views. 
On the other hand, combining URL and uses-feature into one view facilitates formatting uniformity across all views. 
Based on the above considerations, the features are organized into three views, i.e., \textit{Permission View}, \textit{API View}, and \textit{URL \& uses-feature View}. The detailed types of features and views, as well as their descriptions, are shown in Table \ref{tab:View-Feature}.

\begin{table*}
\caption{\label{tab:View-Feature}The types of views and features and their description. }
% \resizebox{\textwidth}{!}{%
\begin{tabular*}{\textwidth}{p{2cm}p{4cm}cp{2.5cm}p{5.3cm}}
\toprule [1.2pt]
\textbf{View type} & \textbf{View description} & \textbf{Feature type} & \textbf{Feature subtype} & \textbf{Feature description}\tabularnewline
\midrule
\multirow{2}{2cm}{Permission View} & \multirow{2}{4cm}{Perspectives on application behavior based on the permissions in the application.} & \multirow{2}{*}{permission} & requested permission & The set of permissions required by the application as declared in the xml file.\tabularnewline
 &  &  & used permission & The set of permissions actually used in the application source code.\tabularnewline
\midrule
\multirow{2}{2cm}{API View} & \multirow{2}{4cm}{Perspectives on application behavior based on the use of sensitive APIs in the application source code.} & \multirow{2}{*}{API} & restricted API & The set of APIs that require specific permissions to be applied.\tabularnewline
 &  &  & suspicious API & Some other sensitive APIs used by the application, which may be related to the access of sensitive information and resources.\tabularnewline
\midrule
\multirow{2}{2cm}{URL \& uses-feature View} & \multirow{2}{4cm}{Perspectives on application behavior based on the uses-features declared in xml file and the URLs coding in the APP's source code.} & URL & URL & URLs found in the source code, some of these addresses might be involved in botnets and thus present in several malware samples.\tabularnewline
 &  & uses-feature & uses-feature & Hardware or software feature requirements registered in the xml file, requiring access to specific hardware clearly has security implications, as the use of certain hardware combinations often reflects potentially malicious behavior.\tabularnewline
\bottomrule [1.2pt]
\end{tabular*}
% }
\end{table*}

To extract these features, we employ Androguard \citep{desnos2018androguard} for decompiling the APK file. 
Then, a static analysis based extractor is developed to automatically identify and extract relevant features from the \textit{AndroidManifest.xml} file and \textit{class.dex} file. 
Notably, we utilize PScout \citep{au2012pscout} to obtain \textit{used permission} and \textit{restricted API}, which compose the mapping relationship.

\subsection{Multi-view text generation}
\label{sec:4.2}

To describe the content within the \textit{Permission View}, \textit{API View}, and \textit{URL \& uses-feature View}, we leverage LLM as a domain expert to generate descriptions and summaries.
This approach not only facilitates thorough reasoning and summarization of potential behaviors from these views, but also delves deeper into the explicit semantics of each view.
It should be noted that the detection performance of our system depends heavily on the text quality of the LLM outputs.
How to guide LLM to fully utilize abilities in order to obtain high quality text possible is an important issue we must consider.
In order to fully utilize and unleash the power of LLM in the Android domain, as well as to ensure it outputs what we need in a uniform and standardized format, meticulous prompt engineering is critical.

\begin{figure}
\centering{}\includegraphics[scale=0.55]{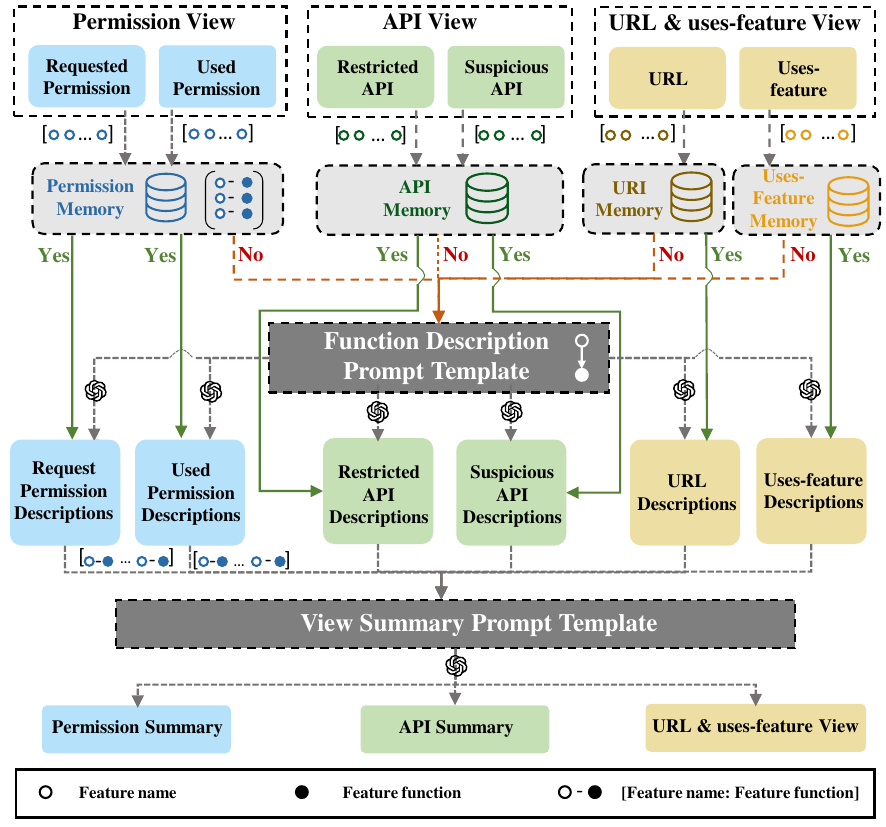}
\caption{The workflow of multi-view prompt engineering guided text generation.}
\label{fig:2}
\end{figure}

In summary, we design a multi-view prompt engineering method as shown in Fig. \ref{fig:2} to generate descriptions and summaries, which fulfill the specific requirements of each view. 
Specifically, the text generation task for each view is divided into two distinct phases: \textit{function description generation} and \textit{view summary generation}. 
For these phases, we designed \textit{function description prompt template} and \textit{view summary prompt template}, enabling the LLM to produce detailed descriptions and summaries for each view.
In this way, we resolve the text generation task into finer-grained phases and views in the form of workflows, ensuring that the LLM generates the appropriate content step by step.
By utilizing this systematic approach, our method achieves impressive detection performance, as demonstrated in Section \ref{sec:5.3}, which also reflects the high quality of the generated text.

It is worth noting that the capabilities of the LLM itself are also an important factor in the quality of the text output.
The LLM employed in our work is \textit{gpt-4-1106-preview} \citep{openai}, which stands out as one of the most well-known and capable model in its field.
In this way, the lower bound on the quality of the text output can be guaranteed.

\subsubsection{Function description generation}
\label{sec:4.2.1}

\textit{Function description} refers to the function explanation of the specific feature under each feature subtype. 
For example, "android.permission.WRITE\_SMS" is the specific feature of the feature type permission, whose \textit{function description} output by LLM is "allow sending and editing SMS". 
The purpose of this sub-module enables the LLM to accurately generate \textit{function description} of each feature, while maintaining our output style via in-context learning.
And then these descriptions are organized into key-value pairs, which are formatted as [Feature name: Feature function], for instance, the aforementioned example is formatted as ["android.permission.WRITE\_SMS": "allows sending and editing SMS"].

For a given APK, the feature extractor extracts the feature string sequence based on different feature subtypes. 
By sequentially injecting each specific feature into \textit{function description prompt template} and subsequently inputting it into the LLM to generate the corresponding \textit{function description}, we obtain a set of feature function key-value pairs for each feature subtype (i.e., \textit{function description list}). This list delineates all specific features and their respective functions within each feature subtype, serving as an intuitive natural language overview of the potential behavior.
The \textit{function description list} on the one hand, is used as one of the important sources for classification detection, and on the other hand, it will be used as an input for obtaining \textit{view summary}, which is a further summary of the behavior of the whole view.

It is imperative to utilize the LLM for generating \textit{function descriptions}, rather than extracting descriptions directly from official documentation \citep{Developer}. 
This necessity arises primarily for two reasons: 1) The official documents provide extensive information on system-level permissions and APIs, making it difficult to automatically extract concise and accurate \textit{function descriptions.} 2) For user-defined permissions, APIs, numerous URLs and other data, the LLM possesses robust summarization capabilities that surpass any alternative automated search and summarization methods. 
For instance, the permission "com.google.android.c2dm.permission.RECEIVE" cannot be directly located and summarized from Android's official documentation. However, through carefully crafted prompt engineering with the LLM, we can obtain its associated \textit{function description}, namely, "allows receiving push notifications from Google Cloud Messaging (GCM)".

Subsequently, we detail the template designs critical for generating \textit{function description}, as illustrated in Table \ref{tab:Example1}.

\begin{table}
\small\sf\centering
\caption{\label{tab:Example1}Patterns of function description prompt template and the generation rule of template.}
\renewcommand\arraystretch{1.3}
\begin{tabularx}{\linewidth}{c|p{1.8cm}|p{5cm}}
\toprule [1.2pt]
\textbf{Id} & \textbf{Prompt pattern} & \textbf{Template of prompt patterns}\tabularnewline
\midrule 
1 & System setup & You are an Android security expert and are familiar with all \{\textit{Feature type}\} and their functions. Please describe the function of the given \{\textit{Feature type}\}:\tabularnewline
\midrule
2 & Example & \{\textit{Feature type}\}: \{\textit{Example feature}\}  \newline function: \{\textit{Function corresponding to feature}\}\tabularnewline
\midrule
3 & Input & Output following the example above. Output only what comes after \textquotedbl function:\textquotedbl . \newline \{\textit{Feature type}\}: \{\textit{Target feature}\} \newline function:\tabularnewline
\midrule
\midrule
\multicolumn{3}{c}{\textbf{Function description prompt generation rule:}}\tabularnewline
\multicolumn{3}{c}{System setup + Example \texttimes $k$ + Input}\tabularnewline
\bottomrule [1.2pt]
\end{tabularx}
\end{table}

\textbf{Function description prompt template. }Given that the \textit{function descriptions} across different feature subtypes exhibit similarities in both format and content, and the generated content is also the essence of features in the form of short sentences, we design template through in-context learning as following.

\begin{itemize}[itemsep=\parskip]

\item \textbf{System setup. }Initially clarify the roles and tasks of LLM through system setup.

\item \textbf{Example. }Subsequently, carefully crafted examples of specific generative patterns and styles are provided for LLM to assimilate. It is worth clarifying that detailed selection and evaluation of the number $k$ of examples is described in Section \ref{sec:5.3}.

\item \textbf{Input. }Finally, clarify the output, i.e., the specific content directly after "function:", and inject a target feature into the template to direct LLM to output the correct content.

\end{itemize}

With the template constructed in the above way, we can guide the LLM to generate \textit{function description} for each target feature and obtain \textit{function description list} for each feature subtype by means of a fixed combination.

\subsubsection{View summary generation}
\label{sec:4.2.2}

In order to explore the deeper potential behavioral information hidden under the features and their functions, we employ LLM to further generate summary for each view by using the \textit{function descriptions list} generated by the aforementioned modules. In this sub-module, we create a step-by-step template to guide the LLM on how to generate \textit{view summaries}. This approach ensures that the LLM can analyze and deduce based on the specified criteria, producing consistently formatted summaries.

\begin{table*}
\small\sf\centering
% \begin{longtable}{c|p{2cm}|p{7cm}|p{7cm}}
\caption{\label{tab:Example2}Patterns of view summary prompt template and the generation rule of template.}
\begin{tabular*}{\textwidth}{c|p{3.5cm}|p{12.5cm}}
\toprule [1.2pt]
\textbf{Id} & \textbf{Prompt pattern} & \textbf{Template of prompt patterns}\tabularnewline
\midrule
1 & System setup & You are an expert in the field of Android security, specializing in auditing Android applications by static analysis. Your task is to combine known information and your expert knowledge to generate a behavior summary for the given Android application in \{\textit{View type}\}.\tabularnewline
\midrule
2 & Task Description & <Task Description>: \newline ``` \newline You must strictly follow the following steps to analyze the application with the package name \textquotedbl\{\textit{Package}\}\textquotedbl
and output a summary from \{\textit{View type}\}: \newline 1- First, you get the \{\textit{Feature type}\}'s contents of the application as follows. The input is in the form of a list, and each element in the list is in the form of '\{\textit{Feature type}\} name: \{\textit{Feature type}\} function':  \newline \hspace*{3mm} 1.1- \{\textit{Feature subtype 1}\}: \{\textit{Function descriptions list 1}\} \newline \hspace*{3mm} 1.2- \{\textit{Feature subtype 2}\}: \{\textit{Function descriptions list 2}\} \newline 2- Now you have known all contents of \{\textit{Feature type}\} of \{\textit{Package}\}. You should start a static analysis from \{\textit{View type}\}, and generate <behavior analysis summary> for the view based on <Output Description and Requirements>. \newline ```\tabularnewline
\midrule
3 & Output Description and Requirements & <Output Description and Requirements>: \newline ``` \newline 1- Output description: Interpretation and summary of known information on \{\textit{View type}\}, focusing on behavior about high-risk \{\textit{Feature type}\} and their potential risks. \newline 2- When you output the summary, do not appeared extra descriptions. Just output the content of the summary. \newline 3- The output must be concise.  \newline 4- Please provide objective summary strictly in terms of \{\textit{View type}\}, and speculation about the behavior of the application should be strictly based on facts and known information. \newline 5- If there is missing information in \{\textit{Feature subtype 1}\} or \{\textit{Feature subtype 2}\}, such as the list is empty, it means that the application has no information about the aspect. \newline 6- Your output should be free of extensions and suggestions, such as \textquotedbl Further exploration is required.\textquotedbl , \textquotedbl Further dynamic analysis is required.\textquotedbl , \textquotedbl Additional information needs to be combined.\textquotedbl etc., as well as your own subjective assumptions, such as \textquotedbl There may be a plausible explanation for these behaviors, but they may also be indicative of potential privacy risks or malicious behaviors of the application.\textquotedbl \newline ``` \tabularnewline
\midrule
4 & Nouns Interpretation & <Nouns Interpretation>: \newline ``` \newline 1- \{\textit{View type}\}: \{\textit{View description}\} \newline 2- \{\textit{Feature type}\}: \newline \hspace*{3mm} 2.1- \{\textit{Feature subtype 1}\}: \{\textit{Feature description 1}\} \newline \hspace*{3mm} 2.2- \{\textit{Feature subtype 2}\}: \{\textit{Feature description 2}\} \newline ```\tabularnewline
\midrule
\midrule
\multicolumn{3}{c}{\textbf{View summary prompt generation rule:} }\tabularnewline
\multicolumn{3}{c}{System setup + Task Description + Output Description and Requirements + Nouns Interpretation}\tabularnewline
\bottomrule [1.2pt]
% \end{longtable}
\end{tabular*}
\end{table*}

\textbf{View summary prompt template. }The design rule of this prompt template are detailed in Table \ref{tab:Example2} and encompass the following four primary components:

\begin{itemize}[itemsep=\parskip]

\item \textbf{System setup. }Initial scoping of LLM role and tasks through appropriate system settings.

\item \textbf{Task description. }Describe to the LLM how to generate the \textit{view summary} step-by-step and introduce the \textit{function descriptions list} to the template so that the LLM can be familiarized with the detailed process of parsing and reasoning.

\item \textbf{Output description and requirement. }Specify the format requirements for LLM output and limit the scope of LLM, which minimizes the output of redundant and worthless information. 
The template also describes countermeasures for the boundary case where the list of feature subtype may be empty for some views of the APK, to prevent LLM from unnecessary hallucinations.

\item \textbf{Nouns interpretation.} Explain to the LLM the meaning of proprietary terms appearing in the template and attempt to guide the LLM in parsing and comprehension.

\end{itemize}

Injecting the pertinent \textit{function descriptions lists} into the prompt structured from the \textit{view summary prompt template} enables the LLM to be guided to generate an appropriate \textit{view summary}.
This facilitates further reasoning and summarization based on the features and their functions to obtain more comprehensive behavioral insights. 
In summary, we provide the LLM with detailed chain-of-thought reasoning, enabling it to summarize the behavior of the views systematically. This approach not only enhances the training of classification models but also helps produce more insightful diagnostic reports.

\subsubsection{Function memory}
\label{sec:4.2.3}

To enhance the generation efficiency and reduce cost (specifically, token consumption) effectively, we introduced memory components within the \textit{function description generation} sub-module to store functions of the four feature types respectively. 
Experience indicates that some system-level permissions, APIs, and other pivotal features frequently play a crucial role in application operations, which means that these features are frequently declared and called by applications. 
As discussed in Section \ref{sec:4.2.1}, generating \textit{function descriptions} primarily involves inserting feature names into designated prompt template before entering them into the LLM. 
This approach may lead to significant token waste due to the repetitive generation of descriptions for certain features, particularly if these features are mechanically fed into the LLM  without any form of memorization.

Based on the preceding discussion, a memory query is performed before invoking LLM to generate a \textit{function description}. 
If a feature name and its corresponding description are already recorded in the database, \textit{function description generation} is bypassed to directly retrieve and compose text matching the formatting requirements. 
Furthermore, for each feature not present in memory, once its \textit{function description} is generated by LLM, we store it in the memory component under the corresponding feature type, thus facilitating the module's efficient operation. 
Experimental results demonstrate that the memory component significantly enhances the module's overall generation efficiency and reduces token consumption, as detailed in Section \ref{sec:5.3}.

\subsubsection{Multi-view prompt implementation}
\label{sec:4.2.4}

Decomposing the text generation task into two sub-phases (i.e., \textit{function description generation} and \textit{view summary generation}) allows us to sequentially connect the entire process in a chain-like manner, yielding higher-quality descriptions and summaries.
Additionally, since our task involves multiple views, it is crucial that the \textit{function description prompt template} and \textit{view summary prompt template} can adapt to this multi-view structure, ensuring that the generated descriptions and summaries align with the diverse content of each view.
Consequently, our multi-view prompt establishes a unified approach to integrating the contents from various views into the templates based on formatting guidelines, facilitating the generation of descriptions and summaries pertinent to each view.
Ultimately, through multi-view prompt engineering, for each view, this module leverages the LLM to generate \textit{function descriptions} and \textit{view summary}. Based on the generated text, detection tasks can be further carried to generate diagnostic reports.

\subsection{Detection classification}
\label{sec:4.3}

For Android malware detection, it is essential to train a classifier that can accurately identify malware. 
It is worth pointing out that our work leverages model concatenation for malware detection to avoid the direct distinction by the LLM based on the descriptions and summaries, with the following reasons: 1) Features derived from static analysis inherently possess limited information, excluding deterministic conclusions based solely on this data, which often requires mining potential patterns from extensive data samples. 
2) Since the LLM is a generalized model, even after rigorous prompt engineering, its direct conclusions about specific domains may still be a hallucination. 
3) The task lacks clear criteria for assessing the magnitude of malicious behavior. 
Supervised training ensures that boundaries are drawn for explicit detection.
Although the generated texts cannot be used directly to determine, they play a crucial role in the generation of the diagnostic report (see Section \ref{sec:4.4} for details).

\begin{table*}
\small\sf\centering
% \begin{longtable}{c|p{2cm}|p{7cm}|p{7cm}}
\caption{\label{tab:Example3}Patterns of diagnostic report prompt template and the generation rule of template.}
\begin{tabular*}{\textwidth}{c|p{3.5cm}|p{12.5cm}}
\toprule [1.2pt]
\textbf{Id} & \textbf{Prompt pattern} & \textbf{Template of prompt patterns}\tabularnewline
\midrule
1 & System setup & You are an expert in the field of Android security, specializing in auditing Android applications by static analysis. Your task is to combine known information and your expert knowledge to generate a diagnostic report for the given Android application.\tabularnewline
\midrule
2 & Task Description & <Task Description>: \newline ``` \newline  You must strictly follow the following steps to analyze the application with the package name "\{\textit{Package}\}" and output a diagnostic report:\newline
1- First, you should know that the application is classified as \{\textit{malicious or benign}\} by the classifier.\newline
2- Then, you get the descriptions and summaries under different views as follows.\newline
    \hspace*{3mm}2.1- <Permission View>\newline
      \hspace*{6mm}2.1.1- <requested permission>: \{\textit{requested permission's function description list}\}\newline
      \hspace*{6mm}2.1.2- <used permission>: \{\textit{used permission's function description list}\}\newline
      \hspace*{6mm}2.1.3- <permission view summary>: \{\textit{permission view summary}\}\newline
    \hspace*{3mm}2.2- <API View>\newline
      \hspace*{6mm}2.2.1- <restricted API>: \{\textit{restricted API's function description list}\}\newline
      \hspace*{6mm}2.2.2- <suspicious API>: \{\textit{suspicious API's function description list}\}\newline
      \hspace*{6mm}2.2.3- <API view summary>: \{\textit{API view summary}\}\newline
    \hspace*{3mm}2.3- <URL \& uses-feature View>\newline
      \hspace*{6mm}2.3.1- <uses-feature>: \{\textit{uses-feature's function description list}\}\newline
      \hspace*{6mm}2.3.2- <URL>: \{\textit{URL's function description list}\}\newline
      \hspace*{6mm}2.3.3- <URL \& uses-feature view summary>: \{\textit{URL \& uses-feature view summary}\}\newline
3- Now you have known not only the application is malicious or not, but also feature function descriptions and view behavior summaries from different views. You should start a static analysis with above information, and generate diagnostic report for the application based on <Output Description and Requirements>.\newline ```\tabularnewline
\midrule
3 & Output Description and Requirements & <Output Description and Requirements>: \newline ``` \newline 1- Your diagnostic report should be based on the above information, focusing on behavior about their potential risks.\newline
2- Your report should contain a summary that describes all possible potential risks in points. The summary must take into account malicious behavior across all views. Each point of potentially malicious behavior needs to point out specific risk points, such as that one feature, API, etc.\newline
3- Your report should provide detailed guidance on next steps for further detection based on summarized potentially malicious behavior.\newline ``` \tabularnewline
\midrule
4 & Nouns Interpretation & <Nouns Interpretation>: \newline ``` \newline \textit{Integrate <Nouns Interpretation> for all views in Table \ref{tab:Example2}.} \newline ```\tabularnewline
\midrule
\midrule
\multicolumn{3}{c}{\textbf{Diagnostic report prompt generation rule:} }\tabularnewline
\multicolumn{3}{c}{System setup + Task Description + Output Description and Requirements + Nouns Interpretation}\tabularnewline
\bottomrule [1.2pt]
% \end{longtable}
\end{tabular*}
\end{table*}

\begin{algorithm}[]\label{alg:1}
	\caption{\textbf{AppPoet} - LLM based Android malware detection via multi-view prompt engineering.}
	\LinesNumbered
	\KwIn{Feature extractor $E$, Permission View $P$, API View $A$, URL \& uses-feature View $U$, function description prompt template $T_{f}$, view summry prompt template $T_s$, diagnostic report prompt template $T_d$, memory component $M$, training data set $D_t$, testing data set $D_e$.}
	\KwOut{The label for the testing Apps $f$, the diagnostic reports for the testing Apps $report$.}
	\For{$view \in \{P, A, U\}$}{
        Get the feature subtype lists $S = \{S_i\}_{i=1}^m$ using $E(view)$;\\
        \For{$i = 1 \to m$}{
            \For{$j = 1 \to \left\vert S_i\right\vert$}{
                \eIf{$S_{ij} \in M$}{
                    Get function $F_{ij}$ of feature $S_{ij}$ from $M$;\\
                }{
                    Generate the function $F_{ij}$ of feature $S_{ij}$ using $T_f(view, S_{ij}) \to \textit{gpt-4-1106-preview}$;\\
                    Put $(S_{ij}, F_{ij}) \to M$; \\
                }
                Put together the function description list $description(S_i)$; \\
            }
        }
        Get the function description lists $description(S)$ of $view$; \\
        Generate $summary$ of $view$ using $T_s(view, description(S)) \to \textit{gpt-4-1106-preview}$;\\
    } 
    Generate representation $Y$ by concatenating $description(P, A, U), summary(P, A, U) \to \textit{text-embedding-ada-002}$;  \\
	Train MLP using $Y_{D_t}$; \\
	\For{$n = 1 \to \left\vert D_e\right\vert$}{
		Generate the label $f_n$ using trained MLP; \\
		Generate the report $report_n$ using  $T_d(f_n, description(P, A, U), summary(P, A, U)) \to \textit{gpt-4-1106-preview}$; \\
	}
	\Return{$f$, $report$.}
\end{algorithm}

To train the classifier, the initial task of this module includes converting the texts from \textit{function descriptions} and \textit{view summaries} generated by the LLM into a machine-readable vector format. 
Considering the need for consistency and recognizing that text embedding is not the focused innovation in this work, we employed OpenAI's embedding model \textit{text-embedding-ada-002} \citep{openai} to transform the generated descriptions and summaries from different views into 1536-dimensional representation vectors. These vectors are then concatenated into a single vector encapsulating the comprehensive behavioral semantic information of the APK from all three views.

Utilizing our dataset and the representation vectors of APKs, a DNN-based classification model is trained, as detailed in Section \ref{sec:5.1} about the selection of the classification model. 
Whenever the detection task of an unknown APK is performed, after obtaining the representation vector of that APK through the aforementioned steps, it can be directly fed into the trained classifier to determine whether the application is malicious or not.

\subsection{Diagnostic report generation}
\label{sec:4.4}

In practical detection environments, merely obtaining a binary detection outcome (malicious or benign) is often insufficient. 
It is crucial to provide a diagnostic report for unknown APKs, which can identify potentially malicious behaviors and guide further investigation or detection. 
Existing learning-based methods still face significant challenges in generating readable and instructive diagnostic reports.

Leveraging the reasoning and summarization capabilities of the LLM, AppPoet is able to generate comprehensible diagnostic reports. By injecting the \textit{function descriptions}, \textit{view summaries} generated by the previous module, and the APK's classification results into the \textit{diagnostic report prompt template} shown in Table \ref{tab:Example3}, AppPoet can produce a diagnostic report for the APK. Since the design of \textit{diagnostic report prompt template} is similar to the \textit{view summary prompt template}, this section does not describe the relevant patterns in the template. In summary, the report can give a comprehensive identification of potential risks and recommendations for next steps in detection based on known information. A specific case is detailed in Section \ref{sec:5.5}. Algorithm. \ref{alg:1} shows the implementation of our developed Android malware detection system AppPoet.

\section{Experiment and evaluation}
\label{sec:5}

In order to verify the detection capabilities of AppPoet, this section aims to explore the following questions:

\begin{itemize}[itemsep=\parskip]

\item \textbf{RQ1: } How does the detection performance of AppPoet in real-world applications compared to that of feature engineering method Drebin and its variant?
\item \textbf{RQ2: } Does the multi-view prompt engineering method designed in AppPoet serve a more effective purpose?
\item \textbf{RQ3: } Are the diagnostic reports generated by AppPoet instructive and valid?

\end{itemize}

Guided by the aforementioned questions, we design and execute a series of experiments utilizing real Android application datasets. This section initially outlines the relevant datasets, configurations, and evaluation metrics for our experiments, and subsequently, the experiments are conducted independently to address the posed questions.

\subsection{Experiment setup}
\label{sec:5.1}

\textbf{Dataset.} To objectively assess the multifaceted performance of AppPoet, our dataset comprises 11,189 benign Apps and 12,128 malicious Apps sourced from AndroZoo \citep{allix2016androzoo}, a dataset collects Apps primarily from official App stores like Google Play and uses VirusTotal \citep{VirusTotal} to determine the nature of each App.

\textbf{Configurations.} Table \ref{tab:Experiment} presents the details of our experimental environment and specific configurations.

\begin{table}
\footnotesize\sf\centering
\caption{\label{tab:Experiment}Experiment configurations.}
\renewcommand\arraystretch{1.2}
\begin{tabular*}{\linewidth}{cc}
\toprule [1.2pt]
Configuration & Model number\tabularnewline
\midrule
CPU & Intel Core i9-13900K\tabularnewline
RAM & 64G\tabularnewline
Operation system & Ubuntu 18.04\tabularnewline
GPU & NVIDIA GeForce RTX 3090 (24G)\tabularnewline
\bottomrule [1.2pt]
\end{tabular*}
\end{table}

\textbf{Evaluation metrics.} The evaluation metrics employed in our experiments are Accuracy, Precision, Recall, and F1-Score, as delineated in Table \ref{tab:Evaluation-metrics}.

\begin{table}
% \small\sf\centering
\caption{\label{tab:Evaluation-metrics}Description of evaluation metrics.}
% \renewcommand\arraystretch{1.3}
% \tabcolsep=0.120\linewidth
\centering{}
\resizebox{0.5\textwidth}{!}{%
\begin{tabular}{cc}
\toprule [1.2pt]
Metrics & Descriptions\tabularnewline
\midrule
$TP$ & The number of correctly identified malicious Apps \\
$TN$ & The number of correctly identified benign Apps \\
$FP$ & The number of misidentified benign Apps \\
$TN$ & The number of misidentified malicious Apps \\
$ACC$ & $(TP+TN)/(TP+TN+FP+FN)$\tabularnewline
$Precision$ & $TP/(TP+FP)$\tabularnewline
$Recall$ & $TP/(TP+FN)$\tabularnewline
$F1$ & $(2\times Precision \times Recall)/(Precision + Recall)$\tabularnewline
\bottomrule [1.2pt]
\end{tabular}
}
\end{table}

\textbf{Classification model selection. }Although the design of classification model is not the main focus of this paper, selecting an appropriate model is crucial for achieving accurate detection results. To this end, we evaluate several common models, including CNN \citep{simonyan2014very}, TextCNN \citep{chen2015convolutional}, RNN \citep{elman1990finding}, LSTM \citep{hochreiter1997long}, and MLP \citep{schmidhuber2015deep}, using the real-world application dataset collected above. This dataset was split into 80\% for training and 20\% for testing. Based on the results in Table \ref{tab:Comparison2}, the MLP model demonstrated the best overall performance. Given its strong ability to preserve the semantic richness of the feature representations, we chose MLP as the classification model for all subsequent experiments.

\begin{table}
\small\sf\centering
\caption{\label{tab:Comparison2}Comparison of different classification model. }
\resizebox{\linewidth}{!}{
\begin{tabular}{ccccc}
\toprule [1.2pt]
Method & ACC(\%) & Precision(\%) & Recall(\%) & F1(\%)\tabularnewline
\midrule
CNN & 95.79 & 96.04 & 95.71 & 95.87\tabularnewline
TextCNN & 96.03 & 96.05 & 96.17 & 96.11\tabularnewline
RNN & 95.94 & 96.92 & 95.08 & 95.99\tabularnewline
LSTM & 96.07 & 96.68 & 95.59 & 96.13\tabularnewline
MLP & 97.15 & 97.03 & 97.39 & 97.21\tabularnewline
\bottomrule [1.2pt] 
\end{tabular}
}
\end{table}

\subsection{RQ1: Performance of AppPoet}
\label{sec:5.2}

To evaluate the detection capability of AppPoet in real-world applications, we compare it with several learning-based methods: (1) Drebin \citep{arp2014drebin} which is String-based method; (2) LBDB \citep{sun2021android} which is Image-based method; and (3) MaMaDroid \citep{onwuzurike2019mamadroid} and Malscan \citep{wu2019malscan} which are both Graph-based methods.
The setup of these baseline methods is explained as follows.
For Drebin, to ensure consistency, we align its input features with the feature subtypes used by AppPoet, as shown in Table \ref{tab:View-Feature}. Additionally, to enhance Drebin's performance, we train an MLP variant of its classification model.
For MaMaDroid, it has two versions that abstract API calls into either family calls or package calls. We configure and conduct experiments on both versions accordingly.
For LBDB and Malscan, we conduct experiments and configurations according to their open-source code.
For AppPoet, we parse the feature subtypes according to the method of multi-view prompt proposed in Section \ref{sec:4.2} and use the method of Section \ref{sec:4.3} to train the classification model. 
We randomly divide 80\% from the real-world application dataset for training and the rest for testing.

\begin{table}[tt]
\small\sf\centering
\caption{\label{tab:Comparison1}Comparison of malware detection performance
for different methods}
\resizebox{\linewidth}{!}{
\begin{tabular}{ccccc}
\toprule [1.2pt]
Method & ACC(\%) & Precision(\%) & Recall(\%) & F1(\%)\tabularnewline
\midrule
Drebin-SVM & 94.76 & 94.60 & 95.33 & 94.96\tabularnewline
Drebin-MLP & 96.06 & 96.47 & 96.04 & 96.26\tabularnewline
LBDB & 91.39 & 92.55 & 88.52 & 91.27\tabularnewline
MaMaDroid-fml & 94.86 & 93.80 & 96.59 & 95.18\tabularnewline
MaMaDroid-pkg & 95.35 & 94.72 & 96.51 & 95.61\tabularnewline
Malscan & 95.65 & 95.54 & 96.15 & 95.84\tabularnewline
AppPoet & 97.15 & 97.03 & 97.39 & 97.21\tabularnewline
\bottomrule [1.2pt]
\end{tabular}
}
\end{table}

Table \ref{tab:Comparison1} presents the detection results of different methods. 
The results show that AppPoet outperforms all other baseline methods across various metrics.
From the results, we can draw the following conclusions: (1) Image-based methods, which directly convert file-level features into images, tend to overlook key semantic information, resulting in the lowest detection performance among the methods. (2) Although String-based methods have limited ability to capture deep semantic information compared to Graph-based methods, their extensive feature extraction allows them to achieve detection performance comparable to that of Graph-based methods. (3) Our method leverages the LLM's strong reasoning and summarization capabilities to further explore the behavioral semantics of the features, leading to superior performance.

\subsection{RQ2: Performance of prompt engineering}
\label{sec:5.3}

Since AppPoet's detection relies on the descriptions and summaries generated by LLM, the quality of these texts significantly influences the representational capabilities of the vectors, and consequently the detection outcomes. 
Therefore, it is crucial to evaluate the effectiveness of the multi-view prompt engineering method proposed in this paper.
In this section, we first perform ablation experiments to assess the importance of different views and text descriptions in influencing detection performance. 
Next, we conduct ablation experiments on workflow design to evaluate the effectiveness of our prompt workflow in generating descriptions and summaries. 
Following this, selection experiments are carried out to determine the optimal number $k$ of examples required for \textit{function description prompt template}, balancing the trade-off between quality and efficiency. 
Lastly, we perform memory and efficiency experiments, where we assess the role of the memory component and evaluate the real-world detection efficiency of AppPoet.

\textbf{Ablation experiments about different views and texts. }To validate the effectiveness and necessity of the various factors in our method, ablation experiments are conducted based on the dataset described in Section \ref{sec:5.1} and the identical experimental setup in Section \ref{sec:5.2}. 
First, we start with three views of AppPoet, eliminating one view at a time, implementing AppPoet-nopermission, AppPoet-noapi, and AppPoet-nourl\&uses-feature, respectively.
Then, to assess the impact of \textit{function descriptions} and \textit{view summaries}, we eliminate the respective view's descriptions and summary, implementing AppPoet-nodescription and AppPoet-nosummary for classification detection. To maintain model consistency, zeros are assigned to the original vector positions upon elimination of specific factors, thereby indirectly fulfilling the ablation objective. The experimental results are presented in Table \ref{tab:Detection-results-of}.

\begin{table}
\small\sf\centering
\caption{\label{tab:Detection-results-of}Detection results after eliminating different views and texts.}
\renewcommand\arraystretch{1.3}
\resizebox{\linewidth}{!}{
\begin{tabular}{ccccc}
\toprule [1.2pt]
Method & ACC(\%) & Precision(\%) & Recall(\%) & F1(\%)\tabularnewline
\midrule
AppPoet-nopermission & 95.43 & 94.83 & 96.30 & 95.56\tabularnewline
AppPoet-noapi & 94.72 & 94.23 & 95.50 & 94.86\tabularnewline
AppPoet-nourl\&uses-feature & 95.92 & 95.78 & 96.26 & 96.02\tabularnewline
AppPoet-nodescription & 95.11 & 94.57 & 95.92 & 95.24\tabularnewline
AppPoet-nosummary & 96.51 & 96.99 & 96.13 & 96.56\tabularnewline
AppPoet & 97.15 & 97.03 & 97.39 & 97.21\tabularnewline
\bottomrule [1.2pt]
\end{tabular}
}
\end{table}

The experimental results indicate that ablating any influencing factor in AppPoet results in a diminished classification performance compared to the original AppPoet, thus strongly affirming the effectiveness of the multi-view \textit{function description} and \textit{view summary}. 
Additionally, the following conclusions can be drawn: 1) From the perspective of views, the API View exerts the most significant impact on the outcomes, whereas the URL \& uses-feature View impacts the results the least, which reflects the varying importance of different views in characterizing malware.
API View is the final link and key indicator for triggering malicious behaviors, which naturally has the greatest impact. 
Moreover, despite the varying importance of the views, combining multiple views indeed enhances the semantic richness of the representation vectors and elevates the detection outcomes. 
2) From the results, utilizing LLM to generate \textit{function descriptions} and further reason and summarize potential behaviors in each view strengthens the capability to mine and represent behavioral information.

\textbf{Ablation experiments about workflow design. }Our multi-view prompt engineering approach is to generate corresponding \textit{function descriptions} and \textit{view summaries} for different views in a \textit{multi-view, multi-phase} manner. 
With this batch-phase design, we can enhance LLM's attention to each detail and generate textual information that is as objective and adequate as possible, which not only ensures the model's detection performance, but also beneficial for improving the richness of diagnostic reports. 
In order to validate the effectiveness of our approach, we designed the prompt template with \textit{multi-view, no phase}, and the prompt template with \textit{multi-phase, no view} (see the Appendix \ref{AppendixA} for details). 
Specifically, the \textit{multi-view, no phase} prompt engineering designs a prompt template for each of the Permission, API and URL \& uses-feature Views. 
After injecting a list of features extracted from the static extractor into it, the LLM generates the \textit{function description} and \textit{view summary} of the view's features directly based on the template's chain guidance. 
\textit{Multi-phase, no view} prompt engineering, in contrast, divides the generation task into two phases, description generation and summary generation, designs a prompt template for each of these phases, and follows the AppPoet approach of in-context learning and thought of chain guidance approach. 
But unlike AppPoet, in this prompt engineering method, we converge all views together and output both the function description and view summary for all three views at once. 
To reduce experimental overhead while ensuring a thorough validation of the method's performance, we randomly select 2,000 malicious samples and 2,000 benign samples from the dataset, forming a balanced subset of 4,000 samples fro small-scale experiments.
Our experiments evaluate the success number of different methods for outputting text and the success number of outputting text that conforms to the required format and can be parsed and embedded by automation. 
Finally, the samples are divided into 80\% training set and 20\% test set and the accuracy of the detection is evaluated by the same MLP model. The experimental results are shown in Table \ref{tab:workflow_design}.

\begin{table*}
\caption{\label{tab:workflow_design}Comparison of different prompt workflows and templates.}

\centering{}
\footnotesize

\renewcommand\arraystretch{1.3}

\resizebox{\textwidth}{!}{%

\begin{tabular}{cccccc}
\hline 
Method & View & $\begin{array}{c}
\text{The number of}\\
\text{successful outputs}\\
(\text{benign | malicious})
\end{array}$ & $\begin{array}{c}
\text{The number of}\\
\text{successful embeddings}\\
\text{(\text{benign | malicious})}
\end{array}$ & ACC $(\%)$ & F1 $(\%)$\tabularnewline
\hline 
\multirow{1}{*}{multi phase, no view} & \multirow{1}{*}{-} & \; 1979 | 1984 & \; 1977 | 1984 & \multirow{1}{*}{93.44} & \multirow{1}{*}{92.25}\tabularnewline
\hline 
\multirow{3}{*}{no phase, multi view} & \multirow{1}{*}{Permission} & \; 1962 | 1989 & \; 1901 | 1938 & \multirow{3}{*}{91.36} & \multirow{3}{*}{91.31}\tabularnewline
 & \multirow{1}{*}{API} & \; 1983 | 1971 & \; 1887 | 1910 &  & \tabularnewline
 & \multirow{1}{*}{URL \& uses-feature} & \; 1990 | 1999 & \; 1989 | 1999 &  & \tabularnewline
\hline 
\multirow{3}{*}{$\begin{array}{c}
\text{multi phase, multi view}\\
(\text{ours})
\end{array}$} & \multirow{1}{*}{Permission} & \; 2000 | 2000 & \; 2000 | 2000 & \multirow{3}{*}{95.50} & \multirow{3}{*}{95.37}\tabularnewline
 & \multirow{1}{*}{API} & \; 2000 | 2000 & \; 2000 | 2000 &  & \tabularnewline
 & \multirow{1}{*}{URL \& uses-feature} & \; 2000 | 2000 & \; 2000 | 2000 &  & \tabularnewline
\hline 
\end{tabular}

}
\end{table*}

As can be seen from the table, the \textit{multi-view, multi-phase} approach adopted by AppPoet outputs all the text successfully, and all the text is formatted, parsed and embedding, which obtains a detection accuracy of 95.50\% of the ACC and 95.37\% of the F1 value. In contrast, neither the multi-view-only, nor the multi-phase-only prompt engineering approach is capable of outputting the full amount of textual information. This is mainly because our experiments are based on utilizing OpenAI's API to communicate with a remote server network. 
In the case of batch calls, if the prompt of a single input is too long or too much output is requested, there is a certain probability that the interaction fails due to the network problem of not being able to deliver the message properly. 
More seriously, the no-phase approach, due to the excessive requirements on the content of a single output from the LLM, can easily lead to the inability of the LLM to output the content in strict accordance with the preset format and requirements, as well as the inability of extracting the effective information to be transformed into embedding vectors for the discriminative process. 
The results of this method have a large amount of output text that cannot be utilized and transformed. 
In addition, the detection performance ultimately obtained by these two methods lags significantly behind that used by AppPoet. In summary, the prompt engineering approach we designed can obtain better detection performance while ensuring correct output and transformation.

\textbf{Selection experiments about the number $k$ of examples.}
In \textit{function description generation} process, we set $k$ examples to guide the LLM in producing more accurate and concise \textit{function descriptions} (as shown in Table \ref{tab:Example1}). 
To determine the optimal value of $k$ that balances effectiveness and efficiency, we conduct the experiments shown in Table \ref{tab:example} based on the 4000 samples randomly selected in the previous dataset.

From the results, we can draw the following conclusions.
(1) In terms of both detection performance and efficiency, providing examples yields better results than not providing them, demonstrating the necessity and superiority of in-context learning.
(2) The results also show that increasing the value of $k$ does not significantly improve detection accuracy but does lead to higher token and time consumption. 
We believe this is because the task of generating \textit{function descriptions} for the relevant features is not particularly challenging for the LLM, a few examples are sufficient to clarify the desired output style and format. 
Based on these findings, we ultimately selected $k=3$ as the optimal number of examples for the \textit{function description prompt template}.

\begin{table*}[tt]
\caption{\label{tab:example}Selection of the number $k$ of examples.}

\centering{}
\footnotesize
\renewcommand\arraystretch{1.3}

\resizebox{\textwidth}{!}{%
\begin{tabular}{cccccccccc}
\hline 
\multirow{3}{*}{The number of examples} & \multirow{3}{*}{Feature type} & \multicolumn{3}{c}{Function description} & \multicolumn{3}{c}{View summary} & \multirow{3}{*}{ACC} & \multirow{3}{*}{F1}\tabularnewline
 &  & \multicolumn{2}{c}{Token consumption} & \multirow{2}{*}{Time consumption} & \multicolumn{2}{c}{Token consumption} & \multirow{2}{*}{Time consumption} &  & \tabularnewline
 &  & Prompt & Response &  & Prompt & Response &  &  & \tabularnewline
\hline 
\multirow{5}{*}{$k=0$} & Permission & 56.75 & 85.74 & 4.10s & 2086.10 & 324.80 & 13.19s & \multirow{5}{*}{93.75\%} & \multirow{5}{*}{93.65\%}\tabularnewline
 & API & 2.78 & 7.32 & 0.71s & 2214.45 & 381.81 & 15.38s &  & \tabularnewline
 & URL & 4.45 & 13.94 & 1.25s & \multirow{2}{*}{805.91} & \multirow{2}{*}{156.39} & \multirow{2}{*}{7.03s} &  & \tabularnewline
 & uses-feature & 0.68 & 1.78 & 0.15s &  &  &  &  & \tabularnewline
 & Total & 64.66 & 108.78 & 6.21s & 5106.46 & 863.00 & 35.6s &  & \tabularnewline
\hline 
\multirow{5}{*}{$k=3$} & Permission & 72.21 & 7.65 & 0.72s & 840.08 & 268.18 & 8.64s & \multirow{5}{*}{95.50\%} & \multirow{5}{*}{95.37\%}\tabularnewline
 & API & 8.07 & 0.33 & 0.07s & 728.81 & 249.39 & 7.93s &  & \tabularnewline
 & URL & 12.72 & 0.66 & 0.11s & \multirow{2}{*}{600.47} & \multirow{2}{*}{149.85} & \multirow{2}{*}{5.79s} &  & \tabularnewline
 & uses-feature & 2.13 & 0.18 & 0.02s &  &  &  &  & \tabularnewline
 & Total & 95.13 & 8.82 & 0.92s & 2169.36 & 667.42 & 22.36s &  & \tabularnewline
\hline 
\multirow{5}{*}{$k=6$} & Permission & 81.61 & 9.34 & 0.91s & 828.99 & 269.29 & 8.61s & \multirow{5}{*}{95.13\%} & \multirow{5}{*}{95.02\%}\tabularnewline
 & API & 10.53 & 0.42 & 0.08s & 732.46 & 252.21 & 8.04s &  & \tabularnewline
 & URL & 15.49 & 0.59 & 0.11s & \multirow{2}{*}{599.36} & \multirow{2}{*}{153.57} & \multirow{2}{*}{5.76s} &  & \tabularnewline
 & uses-feature & 2.73 & 0.17 & 0.04s &  &  &  &  & \tabularnewline
 & Total & 110.36 & 10.52 & 1.14s & 2160.81 & 675.07 & 22.41s &  & \tabularnewline
\hline 
\multirow{5}{*}{$k=9$} & Permission & 98.55 & 8.38 & 1.04s & 835.82 & 268.04 & 8.92s & \multirow{5}{*}{95.88\%} & \multirow{5}{*}{95.71\%}\tabularnewline
 & API & 15.70 & 0.35 & 0.08s & 736.21 & 253.13 & 8.00s &  & \tabularnewline
 & URL & 17.31 & 0.50 & 0.12s & \multirow{2}{*}{599.47} & \multirow{2}{*}{151.33} & \multirow{2}{*}{5.93s} &  & \tabularnewline
 & uses-feature & 3.72 & 0.18 & 0.04s &  &  &  &  & \tabularnewline
 & Total & 135.28 & 9.41 & 1.28s & 2171.5 & 672.5 & 22.85s &  & \tabularnewline
\hline 
\end{tabular}
}
\end{table*}

\begin{figure*}[!t]
\centering
\includegraphics[scale=0.53]{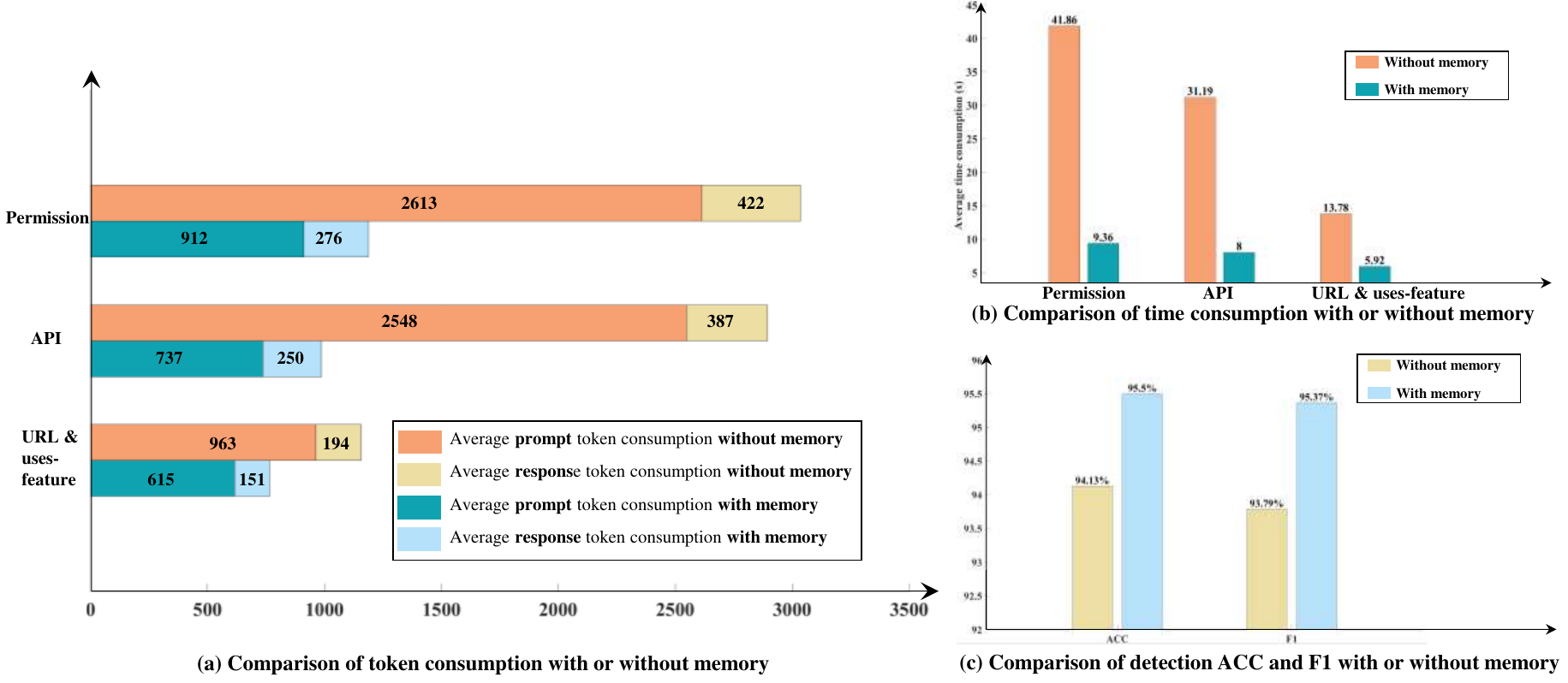}
\caption{Comparison of detection performance with or without memory component.}
\label{fig:table10}
\end{figure*}

\textbf{Memory and efficiency experiments. }In order to evaluate the performance of our designed memory component, we perform a set of comparison experiments. Based on the 4,000 samples randomly selected in the previous dataset, we conduct two sets of experiments with memory and without memory respectively while the multi-view text generator is working, and count the average prompt token consumption and response token consumption of each application in the experiments, as well as the the average time taken to generate text. 
The difference between these two sets of experiments is only whether the memory component is used to store functions. 
The results are shown in Fig \ref{fig:table10}.

As shown in experimental results, our memory component can effectively reduce token consumption and shorten the analysis time to a great extent. 
Thanks to the memory component, we can also guarantee the consistency of the feature function description. 
After manually verifying the functions generated in the experiment without memory, we find that LLM has a certain probability of generating different forms of expressions when generating functions for the same feature. 
For example, the permission "android.permission.ACCESS\_NETWORK\_STATE" is described by LLM as "allow viewing information about network connections" when there is no memory component, but sometimes it is described as "allow checking the state of network connectivity". 
The difference in presentation, though not an obvious wrong, will somewhat affect the presentation of the generated summary and introduce unwanted differences to the embedding representation.
The above situation occurs frequently when there is no memory component, which can be a problem for the classification model to some extent. 
From the ACC and F1-value results of the two sets of experiments, it can be concluded that the detection performance decreases without the memory component.

\begin{table}[tt]
\caption{\label{tab:online}Average detection time consumption.}

\small\sf\centering
\renewcommand\arraystretch{1.3}
\resizebox{\linewidth}{!}{
\begin{tabular}{cc}
\toprule [1.2pt]
Phase of the detection & Average time consumption (s)\tabularnewline
\midrule
Feature extraction & 2.31\tabularnewline
Function description generation & 0.57\tabularnewline
View summary generation & 8.77\tabularnewline
Embedding and detection & 2.37\tabularnewline
Total & 14.02\tabularnewline
\bottomrule [1.2pt]
\end{tabular}
}
\end{table}

To further investigate the real-world efficiency of AppPoet, we train a classifier on the small-scale dataset used in the memory component experiments.
We then introduce 1,000 samples (an equal split between malicious and benign) from the larger dataset as unknown APK inputs. 
Starting from the feature extraction step, we record the average consumption at each stage of the detection process, as shown in Table \ref{tab:online}. 
It is important to note that the memory generated during the \textit{function description generation} phase for the training samples is reused in the process of generating related texts for unknown samples.
Additionally, since the text generation process for each view is entirely independent, we implement a multi-threading approach in the real detection environment to handle different views in parallel.
The results show that AppPoet takes an average of 14.02s to detect a single APK, achieving 93.7\% ACC and 93.64\% F1 score.
While maintaining high detection accuracy, it also demonstrates good efficiency, making it capable of real-time detection and scalability in real-world applications.

\subsection{RQ3: Performance of diagnostic report}
\label{sec:5.5}

Our work exploits the capability of LLM to further mine the behavioral semantic information of an application and obtains excellent detection results, but this is not the end of our work. 
The descriptions and summaries information generated by LLM provides us with a foundational understanding of an application's behavior. 
Combined with the classification outcomes from a high-precision model, a diagnostic report on the application is generated, utilizing LLM's reasoning and summarization capabilities to not only enhance result interpretability but also offer a potent entry point for review and further exploration. 
To verify the validity of the diagnostic report generated by the method outlined in Section \ref{sec:4.4}, a case study is conducted, comparing our report with Drebin's across all aspects, simultaneously affirming the accuracy and validity of our report.

Specifically, a malicious application with the package name "com.applucinante.weddingrings" is randomly selected for analysis, using AppPoet and Drebin \citep{arp2014drebin}. 
Both methods accurately identify the application as malware. 
Then, we generate an interpretable report using the two methods respectively, as shown in Appendix \ref{AppendixB}. It should be noted that we manually verify the application after decompiling it. 
According to the information provided in the report, we locate that the App has a privacy leakage of obtaining sensitive information about the phone and exporting it to outside.

Comparing the two reports we can see that the report generated by Drebin can only be based on the ranking of SVM weights, selecting the top-k highest weighted features that affect the discriminative results as the basis for interpretation, and mechanically assembling them into human-readable statements. However, such a report can provide very little valuable information for researchers. Excluding the factor of misinformation, this report does not allow researchers to have a more comprehensive grasp of the basic information of the application, nor is it very inspiring to give comprehensive reasoning and further ideas for detecting potential malicious behaviors based on known information. With LLM's expert knowledge and linguistic capabilities, AppPoet can take full advantage of all the information available during the detection process to make a comprehensive deduction of the characteristic features and potential behaviors of all the views and give as complete as possible a picture of the possible malicious behaviors. In addition, the interpretability of the feature engineering scheme represented by Drebin mainly comes from the classification models themselves such as SVM and Random Forest. Meanwhile, the weights assigned to key features in their reports are heavily dependent on the selection of datasets, which can also lead to misjudgment to some extent. For example, in the case we provide, the factor that has the greatest impact on determining malicious apps should be "Landroid/telephony/TelephonyManager.getDeviceId", but Drebin scores the weight of this API as 0.20. In contrast, AppPoet takes an unbiased look at all the possible elements of malicious behavior, giving researchers ample inspiration to take the next step.

\section{Discussion}
\label{sec:6}

In this section, we discuss the current limitations in AppPoet as well as the selection and use of LLM. 

\textbf{Current limitations in AppPoet's implementation. }
In Section \ref{sec:4.1}, we completed our work by selecting several features that are more typical of learning-based methods. In fact, our approach is highly scalable and can continuously integrate more static features to combine into different views. In the future, we will continue to expand the richness of the views to enhance the accuracy and generalization of detection. In Section \ref{sec:4.2}, we design rigorous prompt templates to guide the LLM in outputting the specified content according to the requirements and formats. Although we verified the excellent performance of AppPoet in a real application dataset, considering the possibility of LLM's hallucinations and the rigor of the system design, an effective mechanism for checking and correcting errors needs to be established in the next step. In addition, our method is based exclusively on the static analysis of APK internal information, which may introduce a certain amount of false positives in our detection results and diagnostic reports. Therefore, an LLM-driven interactive dynamic and static combined detection method is our next step in this direction. Considering that malware is continuously evolving, we also plan to regularly update and expand the dataset in the future to enhance AppPoet's robustness.

\textbf{The selection and use of LLM. }As mentioned in Section \ref{sec:4.2} and section \ref{sec:4.3}, we select \textit{gpt-4-1106-preview} as the model for text generation and \textit{text-embedding-ada-002} as the embedding model due to their widely proven robustness. However, these models need to be used by way of API calls via network communication. Although our approach obtains the needed information in full volume in small-scale experiments through rational phrase and view disassembly (see Section \ref{sec:5.3} for details), uncontrollable factors such as network fluctuations are still an issue that we need to consider. On the other hand, considering the OpenAI API's feature of billing according to token usage, this approach will continue to incur a significant cost overhead as the training and testing sample sizes continue to increase. As open source LLMs like LLama3 continue to improve in performance, locally deploying and fine-tuning a model with knowledge of the Android security domain can better address the above issues without degrading the quality of the output. 
In addition, in order to control token consumption as much as possible under the premise of guaranteeing the detection performance, we do not really let LLM output their thinking process step by step in the practice of prompt engineering, but rather ensure the output quality by constraining the thinking process of LLM through detailed step-by-step descriptions. 
In the future, we will further design the chain-of-thought process based on open source LLM combined with Retrieval Augmented Generation (RAG) to further ensure the quality and stability of the output.

\section{Conclusion}
\label{sec:7}

With the rapid development of the Android operating system, Android malware detection has emerged as a critical issue within the community. 
Existing String-based and Image-based methods generally lack the mining of semantic information about feature behavior, which affects the detection accuracy of the methods to some extent. 
While graph-based methods are able to mine implicit semantics of features by constructing graphs, they often imply high complexity and abstraction.
Inspired by the success of LLM in natural language understanding, we introduce AppPoet, a novel prompt engineering-based method for Android malware detection.
Specifically, we select permission, API, URL, and uses-feature as entry points for observing Android applications, combining them into three independent views. 
We then direct the LLM to generate function descriptions and view summaries for each view through our proposed multi-view prompt engineering method. 
Subsequently, this textual information is converted into machine-readable representation vectors, enabling malware identification through a trained DNN classifier. 
Finally, we utilize the discrimination results and the descriptive and summary texts of each view to direct the LLM in generating a diagnostic report on the application. This report serves as a guide for reviewing and further analyzing the problem. 
Through the aforementioned method, it becomes possible to further mine behavioral information concealed within the features, to achieve more precise malware detection via the fusion of multi-view information, and to produce a user-friendly diagnostic report.

\section*{Acknowledgements}
This work is supported in part by Anhui Province Natural Science Foundation under Grant No.2408085MF167 and No.2108085QF262, National Natural Science Foundation of China under Grant No.62102385. We thank all the anonymous reviewers who generously contributed their time and efforts. Their professional recommendations have greatly enhanced the quality of the manuscript.

% To print the credit authorship contribution details
% \printcredits

%% Loading bibliography style file
%\bibliographystyle{model1-num-names}
\bibliographystyle{cas-model2-names}

% Loading bibliography database
\bibliography{ref.bib}

% % Biography
% \bio{}
% % Here goes the biography details.
% \endbio

% \bio{pic1}
% % Here goes the biography details.
% \endbio

\appendix
\onecolumn

\section{Prompt templates of other workflow for descriptions and summaries generation}
\label{AppendixA}

\subsection{Multi-view, no-phrase}

You are an expert in the field of Android security, specializing in auditing Android applications by static analysis. Your task is to combine known information and your expert knowledge to generate statements and summaries of objectivity for a given Android application.

\noindent\textbf{<Task Description>:}

```

You must strictly follow the following steps to analyze the application with the package name "\{\textit{Package}\}" and output as <Output Description and Requirements> required:

1- First, you get the \{\textit{View type}\}'s contents of the application as follows:

    1.1- \{\textit{Feature subtype 1}\}: \{\textit{Feature list 1}\}
    
    1.2- \{\textit{Feature subtype 2}\}: \{\textit{Feature list 2}\}
    
2- Now you get all contents of \{\textit{View type}\} of \{\textit{Package}\}. You should start a static analysis from the view, and generate Function description and View summary for the view as <Output Description and Requirements>.

```

\noindent\textbf{<Output Description and Requirements>:}

```

1- Please refer strictly to the following output format and requirement to output the contents of the JSON object, it means you should only output the JSON as follow, without any ohter thing:

Output format:

\{

    \hspace*{3mm}"\{\textit{View type}\}": \{
    
        \hspace*{6mm}"Function description": \{
        
             \hspace*{9mm}"\{\textit{Feature subtype 1}\}": \{\textit{Output format of the feature subtype1.}\}
            
             \hspace*{9mm}"\{\textit{Feature subtype 2}\}": \{\textit{Output format of the feature subtype2.}\}
            
        \hspace*{6mm}\}
        
        \hspace*{6mm}"View summary": \{\textit{Output format of the view summary.}\}
        
    \hspace*{3mm}\}
    
\}

2- Please provide function descriptions and view summaries of the application strictly in terms of the \{\textit{View type}\}, and speculation about the behavior of the application should be strictly based on facts and known information.

3- If there is missing information in some aspects, such as content is empty, it means that the application has no information about the aspects.

4- Your output should be free of extensions and suggestions, such as "Further exploration is required," "Further dynamic analysis is required," "Additional information needs to be combined," etc., as well as your own subjective assumptions, such as "There may be a plausible explanation for these behaviors, but they may also be indicative of potential privacy risks or malicious behaviors of the application."

5- When you output the result, do not appeared extra descriptions such as 'JSON' or ``` etc. Just output the JSON object.

```

\noindent\textbf{<Nouns Interpretation>:}

```

1- \{\textit{View type}\}: \{\textit{View description}\} 

2- \{\textit{Feature type}\}:

\hspace*{3mm}2.1- \{\textit{Feature subtype 1}\}: \{\textit{Feature description 1}\}

\hspace*{3mm}2.2- \{\textit{Feature subtype 2}\}: \{\textit{Feature description 2}\}
    
```

\subsection{Multi-phrase, no-view}

\subsubsection{Function description generation}

You are an Android security expert and are familiar with permission, API, URL, uses-feature and their function. Please describe the function of the given feature:

\# example 1:

permission: android.permission.WRITE\_SMS

function: allow sending and editing SMS

\# example 2:

API: android.telephony.TelephonyManager.getSubscriberId

function: subscriber ID retrieval

\# example 3:

uses-feature: android.hardware.screen.landscape

function: landscape screen orientation support for Android devices

\# example 4:

URL: 360.cn

function: Qihoo 360-related domains (a Chinese internet security company known for antivirus software, web browsers, and mobile application stores)

\# Output following the example above. Output only what comes after "function: ".

\{\textit{Feature type}\}: \{\textit{Feature name}\}

function: \{\textit{Target function}\}

\subsubsection{View summary generation}

You are an expert in the field of Android security, specializing in auditing Android applications by static analysis. Your task is to combine known information and your expert knowledge to generate statements and summaries of objectivity for a given Android application.

\noindent\textbf{<Task Description>:}

```

You must strictly follow the following steps to analyze the application with the package name "\{\textit{Package}\}" and output as <Output Description and Requirements> required:

\hspace*{3mm}1- First, you get the descriptions under different views of the application as follows.

    \hspace*{6mm}1.1- <Permission View>
    
      \hspace*{9mm}1.1.1- <requested permission>: \{\textit{requested permission function descriptions}\}
      
      \hspace*{9mm}1.1.2- <used permission>: \{\textit{used permission function descriptions}\}
      
    \hspace*{6mm}1.2- <API View>
    
      \hspace*{9mm}1.2.1- <restricted API>: \{\textit{restricted API function descriptions}\}
      
      \hspace*{9mm}1.2.2- <suspicious API>: \{\textit{suspicious API function descriptions}\}
      
    \hspace*{6mm}1.3- <URL \& uses-feature View>
    
      \hspace*{9mm}1.3.1- <uses-feature>: \{\textit{uses-feature function function descriptions}\}
      
      \hspace*{9mm}1.3.2- <URL>: \{\textit{URL function function descriptions}\}
      
\hspace*{3mm}2- Now you have known feature function descriptions from different views. You should start a static analysis with above information, and generate <view summary> for each view based on <Output Description and Requirements>.

```

\noindent\textbf{<Output Description and Requirements>:}

```

1- Please refer strictly to the following output format and requirement to output the contents of the JSON object, it means you should only output the JSON as follow, without any ohter thing:

Output format:

\{

    \hspace*{6mm}"Permission View Summary": Interpretation and summary of known information on <Permission View>, focusing on behavioral about high-risk permissions and their potential risks.
    
    \hspace*{6mm}"API View Summary": Interpretation and summary of known information on <API View>, focusing on behavioral about high-risk APIs and their potential risks.
    
    \hspace*{6mm}"URL \& uses-feature View Summary": Interpretation and summary of known information on <URL \& uses-feature View>, focusing on behavioral about high-risk URLs and uses-features and their potential risks. 
    
\}

2- Please provide objective summaries of the application strictly in terms of each view, and speculation about the behavior of the application should be strictly based on facts and known information.

3- If there is missing information in some aspects, such as content is empty, it means that the application has no information about the aspects.

4- Your output should be free of extensions and suggestions, such as "Further exploration is required," "Further dynamic analysis is required," "Additional information needs to be combined," etc., as well as your own subjective assumptions, such as "There may be a plausible explanation for these behaviors, but they may also be indicative of potential privacy risks or malicious behaviors of the application."

5- When you output the result, do not appeared extra descriptions such as 'JSON' or ``` etc. Just output the JSON object.

```

\noindent\textbf{<Nouns Interpretation>:}

```

1- <Permission View>: Perspectives on application behavior based on the permissions in the application.

2- <Permission>: Information about the permissions of the application, including:

    \hspace*{6mm}2.1- <requested permission>: the set of permissions required by the application as declared in the xml file.
    
    \hspace*{6mm}2.2- <used permission>: the set of permissions actually used in the application source code.
    
3- <API view>: Perspectives on application behavior based on the use of sensitive APIs in the application source code.

4- <API>: Sensitive APIs called by the application, including:

    \hspace*{6mm}4.1- <restricted API>: a set of APIs that require specific permissions to be applied.
    
    \hspace*{6mm}4.2- <suspicious API>: some other sensitive APIs used by the application, which may be related to the access of sensitive information and resources.
    
5- <URL \& uses-feature View>: Perspectives on application behavior based on the uses-features declared in xml e and the URLs coding in the APP's source code.

6- <uses-feature>: Hardware or software feature requirements registered in the xml file for a given application, requiring access to specific hardware clearly has security implications, as the use of certain hardware combinations often reflects potentially malicious behavior.

7- <URL>: URLs found in the source code, some of these addresses might be involved in botnets and thus present in several malware samples.

```

\section{Diagnostic report for \textit{com.applucinante.weddingrings}}
\label{AppendixB}

\subsection{Diagnostic report from AppPoet}

\noindent {\large\textbf{Diagnostic Report: Analysis of "com.applucinante.weddingrings" Application}}

{\large\textbf{Summary of Potential Risks:}}

\hspace*{3mm}\textbf{1. Unauthorized Location Tracking:} The app's use of 'android.permission.ACCESS\_FINE\_LOCATION' and 'android.hardware.location.gps' indicates the capability to track the user's precise location, which could be exploited for malicious purposes such as stalking or data profiling without user consent.

\hspace*{3mm}\textbf{2. Privacy Invasion via Camera and Microphone:} Permissions like 'android.permission.CAMERA' and 'android.permission.RECORD\_AUDIO', combined with hardware features 'android.hardware.camera' and \newline'android.hardware.microphone', suggest the app can capture images, videos, and audio, potentially enabling it to spy on users.

\hspace*{3mm}\textbf{3. Access to Sensitive Device Information:} The app's use of 'android.permission.READ\_PHONE\_STATE' and suspicious APIs like 'Landroid/telephony/TelephonyManager.getDeviceId' could allow it to access and misuse sensitive information such as phone numbers, device IDs, and call status.

\hspace*{3mm}\textbf{4. Unauthorized Data Transmission:} The presence of suspicious APIs like 'Lorg/apache/http/client/methods/HttpPost' implies the app may be transmitting data to remote servers, which could include personal user data collected without proper consent.

\hspace*{3mm}\textbf{5. External Storage Manipulation:} With 'android.permission.WRITE\_EXTERNAL\_STORAGE', the app has the ability to write to and potentially read from external storage, posing a risk of unauthorized data access, modification, or deletion.

\hspace*{3mm}\textbf{6. Potential Eavesdropping:} The app's capability to record audio and access running tasks could be used for eavesdropping on user conversations or monitoring user activities.

\hspace*{3mm}\textbf{7. Interaction with Facebook Services:} The app's interaction with '.facebook.com' URLs suggests data sharing with Facebook, which could lead to privacy concerns if the shared data includes sensitive information obtained through the app's permissions.

\hspace*{3mm}\textbf{8. Misuse of System Services and Notifications:} The use of restricted APIs like \newline'android.app.ActivityManager.getRunningTasks' and 'android.app.NotificationManager.notify' could be indicative of the app's ability to manipulate system services and user notifications for malicious intent.

{\large\textbf{Detailed Guidance for Further Detection:}}

\hspace*{3mm}\textbf{1. Code Review:} Conduct a thorough review of the application's source code, focusing on the implementation of the permissions and APIs identified as high-risk. Look for patterns that suggest data is being collected, stored, or transmitted without user consent.

\hspace*{3mm}\textbf{2. Network Traffic Analysis:} Monitor and analyze the network traffic generated by the app to identify any unauthorized data transmission to external servers, especially to the domains associated with Facebook or other unrecognized entities.

\hspace*{3mm}\textbf{3. Behavioral Analysis:} Perform dynamic analysis by running the app in a controlled environment to observe its behavior, particularly how it uses the camera, microphone, and location services in real-time.

\hspace*{3mm}\textbf{4. External Storage Interaction:} Investigate how the app interacts with external storage, checking for unauthorized file access or modifications.

\hspace*{3mm}\textbf{5. Reverse Engineering:} Reverse engineer the app's binaries to understand the deeper functionality that may not be apparent through static analysis alone, such as obfuscated code or hidden payloads.

\hspace*{3mm}\textbf{6. User Consent Verification:} Verify if the app is obtaining explicit user consent before accessing sensitive features and data, in compliance with data protection laws and best practices.

\hspace*{3mm}\textbf{7. Facebook Data Sharing:} Investigate the nature of the app's interaction with Facebook services to ensure that any data sharing is transparent, consensual, and secure.

\hspace*{3mm}\textbf{8. Update Classifier:} Use the findings from the analysis to update the classifier's dataset and improve the detection of similar malicious patterns in other applications.

By following these steps, further detection efforts can be focused on confirming the potential risks identified in this diagnostic report and taking appropriate action to protect users from the malicious behaviors of the "com.applucinante.weddingrings" application.

\subsection{Diagnostic report from Drebin}

\noindent{\large\textbf{Alert}}

The scanned application 'com.applucinante.weddingrings' may contain malicious functionality.

\noindent{\large\textbf{Explanation}}

0.95 Suspicious API calls: Landroid/support/v4/app/ac.getSystemService

\hspace*{3mm}- App uses suspicious API call Landroid/support/v4/app/ac.getSystemService.

0.36 Restricted API calls: android.net.wifi.WifiManager.isWifiEnabled

\hspace*{3mm}- App calls function android.net.wifi.WifiManager.isWifiEnabled to access WIFI\_STATE.   

0.28 Restricted API calls: android.app.ActivityManager.getRunningTasks

\hspace*{3mm}- App calls function android.app.ActivityManager.getRunningTasks to access GET\_TASKS. 

0.25 Hardware features: android.hardware.screen.landscape

\hspace*{3mm}- App uses hardware feature screen.landscape.

0.20 Suspicious API calls: Landroid/telephony/TelephonyManager.getDeviceId

\hspace*{3mm}- App uses suspicious API call Landroid/telephony/TelephonyManager.getDeviceId.

\end{document}